\documentclass[12pt,preprint]{aastex}

\def\lax{{$\mathrel{\hbox{\rlap{\hbox{\lower4pt\hbox{$\sim$}}}\hbox{$<$}}}$}}
\def\gax{{$\mathrel{\hbox{\rlap{\hbox{\lower4pt\hbox{$\sim$}}}\hbox{$>$}}}$}}

\slugcomment{}

\shorttitle{A supermassive binary black hole with triple disks}
\shortauthors{Hayasaki et al.}

\begin{document}


\title{A supermassive binary black hole with triple disks}


\author{Kimitake Hayasaki\altaffilmark{1}}
\affil{Yukawa Institute for Theoretical Physics, Kyoto University, Oiwake-cho, Kitashirakawa, Sakyo-ku, Kyoto 606-8502, Japan}
\email{kimitake@yukawa.kyoto-u.ac.jp}
\author{Shin Mineshige\altaffilmark{1}}
\affil{Yukawa Institute for Theoretical Physics, Kyoto University, Oiwake-cho, Kitashirakawa, Sakyo-ku, Kyoto 606-8502, Japan}

\and

\author{ Luis C.Ho\altaffilmark{2}}
\affil{The Observatories of the Carnegie Institution of Washington, 813 Santa Barbara Street, Pasadena, CA 91101, USA}


\begin{abstract}
Hierarchical structure formation inevitably leads to the formation of 
supermassive binary black holes (BBHs) with a sub-parsec separation in 
galactic nuclei.  However, to date there has been no unambiguous detection of 
such systems.  In an effort to search for potential observational signatures
of supermassive BBHs, we performed high-resolution smoothed particle 
hydrodynamics (SPH) simulations of two black holes in a binary of moderate 
eccentricity surrounded by a circumbinary disk.  Building on our previous work, 
which has shown that gas can periodically transfer from the circumbinary disk 
to the black holes when the binary is on an eccentric orbit, the current 
set of simulations focuses on the formation of the individual accretion 
disks, their evolution and mutual interaction, and the predicted radiative 
signature.  The variation in mass transfer with orbital phase from the 
circumbinary disk induces periodic variations in the light curve 
of the two accretion disks at ultraviolet wavelengths, but not in the 
optical or near-infrared.  Searches for this signal offer a promising 
method to detect supermassive BBHs.
\end{abstract}



\section{Introduction}
\subsection{Scientific Motivation}

Supermassive black holes are common in galactic nuclei and are 
considered to have coevolved with their host galaxies \citep{mag98,fm00,geb00,yutre02,dim05}.
Since galaxies grow through mergers, it is inevitable that 
supermassive binary black holes (BBHs) will form \citep{ebisu91}.
Supermassive BBHs evolves via three stages \citep{bege80,yu02}.  First, each 
of the black holes sinks independently toward the center of the common 
gravitational potential due to dynamical friction with neighboring stars.  
When the separation between the two black holes becomes less than 
$\sim 1\,\rm{pc}$, angular momentum loss by dynamical friction slows down due 
to the loss-cone effect \citep{roos81, makino97, milo01}.  A hard binary forms 
during the second stage, whose orbit slowly shrinks, until the final stage 
where the black holes merge by emitting gravitational radiation.
Recent hydrodynamic simulations of a gas-rich galaxy merger \citep{may07}
show that the interaction between the black holes and the surrounding stars 
and gas rapidly leads to the formation of a BBH. 
After the two black holes have reached the scale of the circumnuclear disk ($\sim 100\,\rm{pc}$), 
it takes less than 1 Myr for them to form a hard binary on scales of $\sim 1$ pc.

Despite the general expectation that BBHs should form naturally during the 
course of galaxy formation and evolution, the currently available 
observational evidence for BBHs have been largely circumstantial and indirect.
These include the detection of X-shaped structures in radio galaxies 
\citep{me02}, double compact cores with a flat radio spectrum \citep{rod06},
and close pairs of AGNs such as that seen in NGC 6240 \citep{komo}.
As these systems evolve, we expect that their black holes would 
eventually form a strongly gravitationally bound binary on sub-parsec 
scales.  Such small-scale binaries have been inferred to be present from 
the quasi-periodic outbursts detected in OJ~287 \citep{sill88} and the 
proper motions seen toward the compact radio core in 3C~66B
\citep{sudou}, but the interpretation of neither case is unambiguous.

In view of the astrophysical importance of BBHs, it would be desirable to 
devise robust methods for detecting them.  \cite{sun97} showed
that when two unequal-mass black holes interact, for instance in the aftermath 
of a minor galaxy merger, the impact of the secondary black hole on the 
accretion disk around the primary black hole may generate periodic flares.  
These authors invoked this model to explain the brightness variations in 
OJ~287.  In supermassive BBHs with extreme mass ratios, the secondary black 
hole could be embedded in the accretion disk of the primary black hole and 
migrate toward it.  The kind of disk-binary interaction offers a promising 
solution to the loss-cone problem \citep{iv99,go00,armi1,armi2}.
\cite{yu01} proposed that BBHs might be detectable from the integrated 
Fe~K$\alpha$ line of two accretion disks, which would give rise to an unusual 
profile with double or multiple peaks.

\subsection{Our Previous Work}

It is likely that a rotating gas disk (i.e., circumbinary disk) forms around 
supermassive BBH during the second stage of a gas-rich galaxy merger.  If so, 
what will happen as a result of the interaction between the BBH and the 
circumbinary disk?  \cite{haya07}, using a smoothed particle hydrodynamics 
(SPH) code \citep{benz90a,benz90b,bate95}, investigated the accretion flow 
from a circumbinary disk onto the supermassive BBH on sub-parsec scales.  In 
that study, the circumbinary disk was aligned with the binary orbital plane.  
They found that mass transfer can occur from the circumbinary disk to the 
supermassive BBH in two steps.  First, the initially circularized circumbinary 
disk becomes elongated due to the azimuthal $m=2$ component of the binary 
potential. The gas density then grows at the two points closest from the black 
holes. Next, when the gas reaches beyond the potential barrier of the binary, 
mass inflow is initiated by gas pressure so that the gas freely inspirals onto 
each of the black holes.  This process repeats every binary orbit.  The 
viscosity of the circumbinary disk is ineffective for the mass transfer 
process because the viscous timescale is much longer than the orbital period. 
For BBH systems on sub-parsec scales, the viscous timescale of the 
circumbinary disk can be written as
\begin{eqnarray}
\frac{\tau_{\rm{vis}}^{\rm{cbd}}}{P_{\rm{orb}}}
\sim
7.31
\times
10^{5}\left(\frac{0.1}{\alpha_{\rm{SS}}}\right)\left(\frac{10^{4}~\rm{K}}{T}\right)
\left(\frac{M_{\rm{BH}}}{10^{8}~M_{\odot}}\right)\left(\frac{0.01~\rm{pc}}{a}\right), 
\label{eq:tcbd}
\end{eqnarray}
where $\alpha_{\rm{SS}}=0.1$ is the Shakura \& Sunyaev (1973) viscosity 
parameter, $M_{\rm{BH}}$ is the total black hole mass, and $a$ is the 
semi-major axis, respectively.  The inner edge of the circumbinary disk is 
assumed to be at the $1:3$ resonance ($\sim2.08\,a$), where the viscous force 
is balanced with the tidal/resonant force of the binary \citep{al94}. 
The circumbinary disk is assumed to follow an isothermal equation of state.

The mass transfer rate significantly depends on the binary orbital phase in 
eccentric binaries, whereas it shows little variation with orbital phase in 
circular binaries.  In both cases the supermassive BBH system is expected to 
have triple disks because the mass transfer from the circumbinary disk would 
inevitably lead to the formation of an accretion disk around each black 
hole.  However, there is little known about how the accretion disks are 
formed and what signatures they exhibit.  We have, therefore, performed a new 
set of simulations at higher resolution, adopting the same set of binary 
orbital parameters we had previously used except for a semi-mjor axis 
$a=0.01\,\rm{pc}$, which was chosen to yield a shorter period binary than that 
of the previous simulation ($a=0.1\,\rm{pc}$; Hayasaki et al. 2007).

The simulations should ideally take into account all of the processes at work 
in a triple-disk system, including mass transfer from the circumbinary disk to 
the individual black holes, and the accretion flow onto each black hole.  Such 
a simulation, however, would require an enormous amount of computational time.  
Therefore, we divide these processes into a two-stage simulation.  In the 
first stage, we investigate how the mass is transferred from the circumbinary 
disk to the effective gravitational radius of each black hole.  In the second 
stage, we confine ourselves to simulate only the accretion onto each black 
hole by setting the mass-transfer rate from the first-stage simulation as a 
new boundary condition.  In this paper, we focus on the results of the 
second-stage simulation.


\section{Numerical Procedures}

We set the binary orbit on the $x-y$ plane with the semi-major axis along the 
$x$-axis. Initially, each black hole is at the periastron.  It orbits with 
$P_{\rm{orb}}\approx 9.4\,\rm{yr}$, and the orbital eccentricity is chosen to 
be $e=0.5$.  As a result of the interaction between the black holes and the 
neighboring stars and gas \citep{may07}, the BBH likely forms rapidly on 
scales of a few parsecs.  The total black hole mass is $10^{8}M_{\odot}$, 
which is a typical value for the centers of massive elliptical galaxies.
The simulations were carried out with a total of 152,000 particles at the end, 
each one having a mass of $1.0\times10^{-15}\,M_{\rm BH}$.
For simplicity, we set the mass ratio to $q \equiv M_2/M_1=1.0$ and assume that 
the circumbinary disk is coplanar with the binary orbital plane.  The effect 
of the mass ratio, the orbital eccentricity, and inclination angle on the mass 
transfer process will be studied in a subsequent paper.  In this section, we 
first explain the boundary conditions given by the first-stage simulation, 
followed by a brief description of the energy equation.

\subsection{Boundary Conditions}

To reduce computational time and to significantly improve the resolution, 
we confine the simulations only to the accretion flows around the supermassive 
BBH by supplying mass periodically from the outer boundary conditions, which are
defined by the SPH particles captured by the black holes \citep{haya04, haya07}.
The capture radii correspond to the effective gravitational radii 
$0.8\,{r_{\rm{L}}}$, where $r_{\rm{L}}$ is the innermost common gravitational 
radius of the binary potential for a circular binary.  The mass is added at 
a rate of 1 $M_\odot\,{\rm yr}^{-1}$ on the outer edge of the circumbinary 
disk.  The system reaches quasi-steady state equilibrium by 30 binary orbits, 
when the mass input balances the mass output.  The mass of the circumbinary 
disk is $\sim10^{-6}\,M_{\rm{BH}}$, and there are $56,870$ SPH particles at 
the end of the simulation.  To reduce the fluctuation noise, the data are 
folded over the orbital period $30\le{t}\le60$, where the unit of time, unless 
otherwise noted, is $P_{\rm{orb}}\simeq9.4\,\rm{yr}$.  The orbital phase 
dependence of the mass transfer rate obtained by this procedure is shown by 
the solid line in Figure~1.  We model the mass transfer process from the 
circumbinary disk to the supermassive BBH by injecting gas particles at a 
given phase-dependent rate with spatial and velocity Gaussian distributions 
derived from the averaged values over $30\le{t}\le60$.  

\subsection{Energy Equation}
The new simulations no longer assume an isothermal equation of state.  We 
introduce an energy equation in order to resolve the temperature profile.
The specific internal energy of an ideal gas is explicitly given by
\begin{equation}
u=\frac{K}{\gamma-1}\rho^{\gamma-1},
\label{eq:2}
\end{equation}
where $\rho$ is the density, $\gamma$ is the adiabatic index for an ideal gas 
(5/3), and $K$ is a function of the entropy, which is obtained by the first 
law of thermodynamics.  Time differentiation of equation~(\ref{eq:2}) yields
the energy equation, which can be written as
\begin{equation}
\frac{du}{dt}=
\frac{P}{\rho^{2}}\frac{d\rho}{dt} + 
\frac{\rho^{\gamma-1}}{\gamma-1}\frac{dK}{dt}.
\label{eq:3}
\end{equation}
The first term on the right-hand side gives the heating of the gas at 
pressure $P$. In the current simulation, we assume that the heating due to 
viscosity is dissipated by cooling due to blackbody radiation. The cooling 
term is related to the effective temperature $T$ by the Stefan-Boltzmann law,
\begin{equation}
\frac{2\sigma{T^{4}}}{\rho{H_{\rm{c}}}}=\frac{\rho^{\gamma-1}}{\gamma-1}\frac{dK}{dt},
\label{eq:4}
\end{equation}
where $\sigma$ is the Stefan-Boltzmann constant and $H_{\rm{c}}$ is the characteristic scale of the system, 
and the factor 2 represents radiation from the two sides of the accretion disk.

In SPH calculations, gas heating is implemented using a standard
SPH artificial viscosity \citep{mona83}
\begin{eqnarray}
\Pi_{ij}=
\left\{
\begin{array}{ll}
(-\alpha_{\rm{SPH}}c_{\rm{s}}\mu_{ij} + \beta_{\rm{SPH}}\mu_{ij}^{2})
/\rho_{ij} 
& \mbox{{\rm for}\, \boldmath $v$}_{ij}\cdot\mbox{\boldmath $r$}_{ij}\le0 \\
0
& \mbox{{\rm for}\, \boldmath $v$}_{ij}\cdot\mbox{\boldmath $r$}_{ij}>0 \\
\end{array}
\right.
\label{eq:5}
\end{eqnarray}
for particle $i$ and its neighboring particle $j$, where
$\alpha_{\rm{SPH}}$ and $\beta_{\rm{SPH}}$ are the linear and
nonlinear artificial viscosity parameters, respectively,
$\rho_{ij}=(\rho_{i}+\rho_{j})/2$ is the mean density, 
$\mbox{\boldmath $r$}_{ij}=\mbox{\boldmath $r$}_{i}
-\mbox{\boldmath $r$}_{j}$ is the relative position vector, 
$\mbox{\boldmath $v$}_{ij}=\mbox{\boldmath $v$}_{i}
-\mbox{\boldmath $v$}_{j}$ is the relative velocity vector, 
and $\mu_{ij}=h_{ij}\mbox{\boldmath $v$}_{ij}\cdot\mbox{\boldmath $r$}_{ij}
/(r_{ij}+\eta_{ij})$ with $\eta_{ij}^{2}=0.01h_{ij}^{2}$ and 
$h_{ij}=(h_{i}+h_{j})/2$ being the mean smoothing length.
The entropy change of particle $i$ can be written as \citep{bate295}
\begin{equation}
\frac{dK_{i}}{dt} = \frac{\gamma-1}{2\rho_{i}^{\gamma-1}}\sum_{j}^{N_{\rm{n}}}m_{j}\Pi_{ij}
\mbox{\boldmath $v$}_{ij}
\cdot
\mbox{\boldmath $\nabla$}_{i}W(\mbox{\boldmath $r$}_{ij},h_{ij}),
\label{eq:6}
\end{equation}
where $N_{\rm{n}}$ is the number of neighboring particles and $W$, 
proportional to $|h_{ij}|^{-3}$, is called the SPH kernel.  Thus, the effective 
temperature of particle $i$ can be calculated from 
equations~(\ref{eq:4})-(\ref{eq:6}):
\begin{equation}
T_{i}=\left[
\frac{\rho_{i}}{4\sigma}\sum_{j}^{N_{\rm{n}}}m_{j}h_{ij}\Pi_{ij}
\mbox{\boldmath $v$}_{ij}
\cdot
\mbox{\boldmath $\nabla$}_{i}W(\mbox{\boldmath $r$}_{ij},h_{ij})
\right]^{1/4},
\label{eq:7}
\end{equation}
where we set $H_{c}=h_{ij}$ in the current simulation. 

\section{Accretion Disks around the Black Holes}

Each of the two black holes has an accretion radius $5\times10^{-3}a$, which 
is the radius of the inner boundary of the simulation and corresponds to 
$r_{\rm{in}} \simeq 10.4\, r_{\rm{BH}}$, where $r_{\rm{BH}}$ is the 
Schwarzschild radius of a black hole with mass $5.0\times10^{7}~M_{\odot}$. 
To verify the robustness of our simulation results, we have compared the 
averaged mass accretion rate during one orbital period with that of a 
simulation whereby the number of injected particles was doubled and the 
particle mass was reduced in half.  We find no qualitative or quantitative 
differences to within $\sim8\%$. 

\subsection{Formation Process}

We use the standard SPH artificial viscosity throughout the simulations.  In 
simulating the disks, the shear component of the SPH artificial viscosity can 
be replaced with the Shakura-Sunyaev viscosity parameter $\alpha_{\rm{SS}}$ in 
the following form: $\alpha_{\rm{SS}} = 0.1\alpha_{\rm{SPH}} h / H$ 
\citep{haya07}, where $h$ is the azimuthally averaged  smoothing length and 
$H$ is the azimuthally averaged scale height of the disk. 
The estimated value of $\alpha_{\rm{SS}}$ ranges from $0.001$ to $0.2$ 
over the entire region of the disk.  Little gas can accrete viscously onto the 
black holes within the simulation run time because the viscous timescale is 
much longer than the orbital period (cf. Eq.~\ref{eq:tad}). 

The injected particles originally have eccentric orbits around the black holes 
because their angular momenta are smaller than that of the particles at the 
disk outer edge ($0.2\,a$).  Such gas particles accrete directly onto the black 
holes within one dynamical timescale of the disk, 
$\tau_{\rm{dyn}}=2\pi/\Omega_{\rm{disk}}\sim0.1P_{\rm{orb}}$, which is 
estimated at the disk outer edge.  In addition, the particles in each disk 
lose angular momentum mainly from tidal interaction with the other black hole 
because the timescale for angular momentum loss at the outer edge of the disk
is shorter than the binary orbital period: $\tau_{\rm{tide}} \sim 
2\pi/(\Omega_{\rm{disk}} - \Omega_{\rm{orb}}) \sim 0.1P_{\rm{orb}}$, where 
$\Omega_{\rm{disk}}$ and $\Omega_{\rm{orb}}$ are the angular frequency of the 
disk and of the binary, respectively.  This also makes it possible for the gas 
to accrete onto the black holes within the timescale of the binary orbit. 

We illustrate the behavior of the two accretion disks in Figure~2, which gives 
snapshots of the accretion flow around the supermassive BBH covering one 
entire orbital period during the 10th binary orbit.  At the end of the 
simulation, the disk has a mass of $\sim10^{-8}\,M_{\rm{BH}}$.  The two 
white crosses on each density map indicate the mass supply points.  When the 
two black holes are closest together ($t=10.0$), the disks are deformed due to 
the tidal interaction, and the supplied gas is added at the outer parts of the 
disks ($t=10.04$).  At $t=10.1$, the gas particles that are added at the disk 
outer edge start to inspiral onto the black holes within a dynamical time.  
The disk size become $\sim50\%$ smaller than the disk outer edge.  As the 
black holes recede toward apastron ($t=10.18$), the disks are restored from 
their tidal deformation and gradually circularize; tidal tails connect to the 
mass supply points between the two black holes.  As can be seen in 
Figure~3{\it a}, during this time the temperature of the inner part of the 
disk is highest.  The disks are connected to each other via tidal tails and 
are still eccentric, even at apastron ($t=10.5$).  The disks approach each 
other again toward the next periastron, and the injected particles once again 
start to be added to the disks ($t=10.75$).  This pattern repeats periodically.

\subsection{Survival}

Whether the accretion disks formed around the black holes are transient
or persistent depends on the viscous timescale of the disk compared with the
orbital period.  For simplicity, we assume the accretion disk to be
geometrically thin and isothermal with a Shakura-Sunyaev viscosity parameter
$\alpha_{\rm{SS}}$.  The ratio of the viscous timescale of a disk of size
$R_{\rm{d}}$ to the orbital period is then given by
\begin{equation}
\frac{\tau_{\rm{vis}}^{\rm{bhdisk}}}{P_{\rm{orb}}}
\sim
5.1\times10^{5}\frac{1}{\sqrt{1+q}}\left(\frac{R_{\rm{d}}}{a}\right)^{1/2}
\left(\frac{0.1}{\alpha_{\rm{SS}}}\right)\left(\frac{10^{4}K}{T}\right)\left(\frac{0.01\,\rm{pc}}{a}\right)\left(\frac{M_{\rm{BH}}}{10^{8}M_{\odot}}\right).
\label{eq:tad}
\end{equation}
It is immediately seen that the viscous timescale is much longer than the
binary period.  Thus, the accretion disks will survive, not only over the
whole orbital phase but also over the simulation run-time
($\sim10\,P_{\rm{orb}}$) once they are formed.

If the separation between the two black holes is less than $0.01\,\rm{pc}$,
emission of gravitational waves becomes the dominant mechanism to dissipate
the binary's orbital energy \citep{bege80}.  Dissipation makes the binary
evolve rapidly toward the coalescence of the two black holes.  Therefore, we
should consider whether the accretion disks survive with shrinkage of the
semi-major axis due to gravitational wave emission.  The merging timescale
between two black holes due to the emission of gravitational radiation
is given by \citep{peters64}
\begin{equation}
\frac{\tau_{\rm{gr}}}{P_{\rm{orb}}}\sim6\times10^{5}\frac{(1+q)^{2}}{q}f(e)\left(\frac{0.01\rm{pc}}{a}\right)^{-5/2}\left(\frac{M_{\rm{BH}}}{10^{8}M_{\odot}}\right)^{-5/2},
\label{eq:tgr}
\end{equation}
where $f(e)=(1-e^2)^{7/2}/(1+73e^{2}/24+37e^{4}/96)$.  From equations (7) and
(8), we can compare the viscous timescale of the disk with the merging
timescale:
\begin{equation}
\frac{\tau_{\rm{vis}}^{\rm{bhdisk}}}{\tau_{\rm{gr}}}\sim\frac{q}{(1+q)^{5/2}}\frac{1}{f(e)}\left(\frac{R_{\rm{d}}}{a}\right)^{1/2}
\left(\frac{0.1}{\alpha_{\rm{SS}}}\right)\left(\frac{10^{4}K}{T}\right)\left(\frac{0.01\,\rm{pc}}{a}\right)^{7/2}\left(\frac{M_{\rm{BH}}}{10^{8}M_{\odot}}\right)^{7/2}.
\label{eq:tadgr}
\end{equation}
The range of our simulation parameters satisfy
$\tau_{\rm{gr}}\ge\tau_{\rm{vis}}^{\rm{bhdisk}}\gg P_{\rm{orb}}$.
This ensures that the accretion disks, once formed, would survive until the black holes coalescence 
if the mass supply from the circumbinary disk to the central binary can be maintained.
Gravitational wave emission has little influence on the formation and 
evolution of the accretion disks.

\subsection{Temperature Profiles, Spectral Energy Distributions and Light Curves}

The radial distribution of the azimuthally averaged temperature is shown in 
Figure~3{\it a}.  The solid and dotted lines denote, respectively, the highest 
and lowest temperatures at $r_{\rm{in}}$.  The gas is distributed over the 
entire region of the disk, and the temperature profile varies with orbital 
phase.  Figure~3{\it b}\ shows the corresponding spectral energy distributions 
(SEDs) normalized by the Eddington luminosity for a total black hole mass 
$M_{\rm{BH}}=1.0\times10^{8}~M_{\odot}$.  The SEDs are computed numerically by 
integrating the spectrum, which is a function of disk temperature, from the 
outer edge of the disk ($0.2\,a$) to $r_{\rm{in}}$. Each SED has a peak in the 
ultraviolet (UV) region.

Figure~4{\it a} shows the orbital phase dependence of the light curves emitted 
from the disk in the UV, optical, and near-infrared (NIR) bands, obtained by 
averaging out the time variations over the orbital periods $10\le{t}\le11$.  
While the UV light curve shows substantial fluctuations with time, the optical 
and NIR light curves exhibit little variation. This is because a small 
temperature change produces large-amplitude variations in the UV region, where 
the SED decays exponentially (Fig.~3{\it b}), whereas it has little effect on 
the Rayleigh-Jeans part of the spectrum (optical and NIR).  What does the 
X-ray light curve look like?  The dashed line in Figure~1 shows the orbital 
variation of the mass accretion rate, which is defined as the number of SPH 
particles per unit time captured at $r_{\rm{in}}$ over the orbital period 
$10\le{t}\le11$.  If the gas in the disk accretes onto the black hole while 
keeping this orbital phase dependence, the X-ray light variation is expected 
to track the variation in mass accretion rate. This situation seems possible, 
if the accretion timescale in the inner region is shorter than the binary 
orbital period (i.e., close to the free-fall timescale), as is the case for 
radiatively inefficient accretion flows \citep{kato9807}.  The recent 
hydrodynamical simulations of supermassive BBHs by \cite{bog08} predict 
strong X-ray outbursts during periastron passage.  However, since our 
simulations do not take into account the relativistic correction of the 
gravitational potential in our simulations, the detail properties of the X-ray 
emission are beyond the scope of this work.

In order to isolate the effect of the mass transfer from the circumbinary disk 
onto the accretion disks, we carried out a simulation in which the mass 
transfer was artificially stopped over the orbital 
period $11\le{t}\le12$.  For comparison purpose, the light curves for this 
case are shown in Figure~4{\it b}.  Note that none of the light curves varies 
much with orbital phase.  Thus, the periodic mass transfer from the 
circumbinary disk is clearly responsible for the significant variation of the 
UV light curve.


\section{Summary and Discussion}

We have performed SPH simulations of accretion flows around each black hole in 
a supermassive BBH system, taking into account periodic mass transfer from 
the circumbinary disk.  An accretion disk forms and survives around each black 
hole.  Thus, supermassive BBHs have triple disks, consisting of two accretion 
disks, one around each black hole, and a circumbinary disk surrounding them as 
a mass reservoir.  In such systems, three dynamical processes are expected to 
give rise to structures in the two accretion disks.  First, periodic mass 
transfer makes the disks eccentric because the gas particles from the 
circumbinary disk originally have elliptical orbits around the black holes.  
This makes it possible to accrete directly onto the black holes during the 
binary orbit.  Second, once the disks form, periodic mass transfer from the 
circumbinary disk episodically gives an impact to the outer edge of the disks, 
resulting in the excitation of a one-armed spiral wave in the disks (e.g., 
Hayasaki \& Okazaki 2005).  Mass exchange between the two disks further 
reinforces the one-armed spiral wave.  Finally, the disks are tidally deformed 
by the time-varying binary potential owing to the orbital eccentricity. 
 
During the simulation, the accretion disks are still developing because their 
viscous timescale is much longer than the simulation run time 
($10\,P_{\rm{orb}}$).  In addition, a one-armed spiral wave could also 
contribute to the non-axisymmetric disk structure because its propagation 
timescale, $\tau_{\rm{wave}}\sim(\alpha_{\rm{SS}}/2\pi)\tau_{\rm{vis}}^{\rm{bhdisk}}\sim10^3P_{\rm{orb}}$, is also much longer than the simulation run time \citep{haya05}.  
In such an early stage of the disk evolution, the periodic mass transfer has 
the dominant effect on the disk structure.  Without mass transfer, tidal 
deformation alone has little influence on the light curve variation.  As a 
consequence of the periodic mass transfer, we show that the light curve in the 
UV --- and mostly likely also in the X-rays --- exhibits variations with 
orbital phase, whereas little fluctuation is seen in the optical and NIR light 
curves.  NIR variability might be detectable if the binary is surrounded by 
a dusty torus, whose inner edge would be irradiated by, and thus mirror, the 
variable central UV and X-ray sources.  The photometric variability discussed
in this paper is distinctive from the intrinsic variability normally seen from 
single accretion disks or jets, and thus provides a unique signature to 
identify BBH systems in galactic nuclei.

The orbital period of the fiducial case simulated here, $\sim 9.4$ yr, is long 
for any realistic observational experiment.  The only known BBH candidate 
with a similar orbital period (11--12 yr) is the blazar OJ\,287 \citep{sill88}.
Supermassive BBHs with shorter orbital periods (e.g., up to a few years) would 
be more practical observationally.  However, the generic features of the
triple-disk BBH system we have simulated would remain unchanged for shorter 
orbital periods.  To avoid general relativistic corrections to the potential, 
the size of the accretion disk needs to be larger than $\sim30\, r_{\rm{BH}}$. 
Since the semi-major axis of an equal-mass binary is \gax\ 20 times the disk 
size, the results of our simulations can be scaled to systems with $a$ \gax\ 
600 $r_{\rm{BH}}$.  This condition gives a wide range of tractable orbital 
periods, from $P_{\rm{orb}}=5\,\rm{d}$ for $M_{\rm{BH}}=10^{6}\,M_{\odot}$ to 
$P_{\rm{orb}}=1.4\,\rm{yr}$ for $M_{\rm{BH}}=10^{8}\,M_{\odot}$. 

In the first-stage simulation, the orbital-phase dependence 
of mass transfer rate is smoothed out 
as the orbital eccentricity gets toward zero.  On the other hand, 
the peak of the mass transfer rate in a higher eccentricity binary is sharper
(cf. \citealt{haya07}).
These features can also be seen in the mass accretion rates 
and the light curves in the second-stage simulation. 
Most of the angular momentum gained from the mass transfer 
from the circumbinary disk is converted into that of the BBH 
via the tidal interaction between the BBH and the accretion disk. 
These processes damp the orbital eccentricity of the binary. 
However, since the damping timescale approximately equals the viscous timescale, 
which is much longer than the binary orbital period (cf. equation~\ref{eq:tad}), 
this effect cannot be seen in this simulation.

When the separation between two black holes falls below $\sim 0.01\, \rm{pc}$, 
the binary will suffer rapid evolution due to the emission of 
gravitational radiation \citep{bege80}.  With the shrinkage of the semi-major 
axis, the gap between the circumbinary disk and the binary would be reformed.
The timescale for gap reformation can be estimated as 
$\tau_{\rm{gap}}\sim10^{-4}\tau_{\rm{vis}}^{\rm{cbd}}$ using equation~(\ref{eq:tcbd}) 
(cf. \citealt{al94}).  If $\tau_{\rm{gr}}<\tau_{\rm{gap}}$, mass transfer 
from the circumbinary disk will be stopped, and the light curves from the 
system would exhibit little periodicity as in Figure~4{\it b}.  On the other 
hand, if $\tau_{\rm{gr}}>\tau_{\rm{gap}}$, mass will continue to be supplied 
from the circumbinary disk to each black hole, and the BBH system will maintain
triple disks.  In this case, the periodic light curves predicted for the 
accretion disks can serve as the electromagnetic counterparts to gravitational 
wave events detectable with instruments such as {\it Laser Interferometer 
Space Antenna}.

\acknowledgments

We thank an anonymous referee for many useful comments and suggestions.
K.H. is grateful to Atsuo~T. Okazaki, Daisuke Kawata, and Hiroshi Sudou 
for helpful discussions.  The authors thank YITP in Kyoto University, where 
this work was extensively discussed during the workshops YITP-W-05-11 on 
September 20--21, 2005 and YITP-W-06-20 on February 13--15, 2007.  The 
simulations reported here were performed using the facility at the Centre for 
Astrophysics \& Supercomputing at Swinburne University of Technology, 
Australia and at YITP in Kyoto University.  This work was supported in 
part by the Grants-in-Aid of the Ministry of Education, Science, Culture, and 
Sport and Technology (MEXT; 30374218 K.H., 14079205 K.H. \& S.M.), 
and by the Grant-in-Aid for the 21st Century COE Scientific Research Programs 
on ``Topological Science and Technology'' and ``Center for Diversity 
and Universality in Physics'' from MEXT.


\clearpage

\begin{figure*}
\plotone{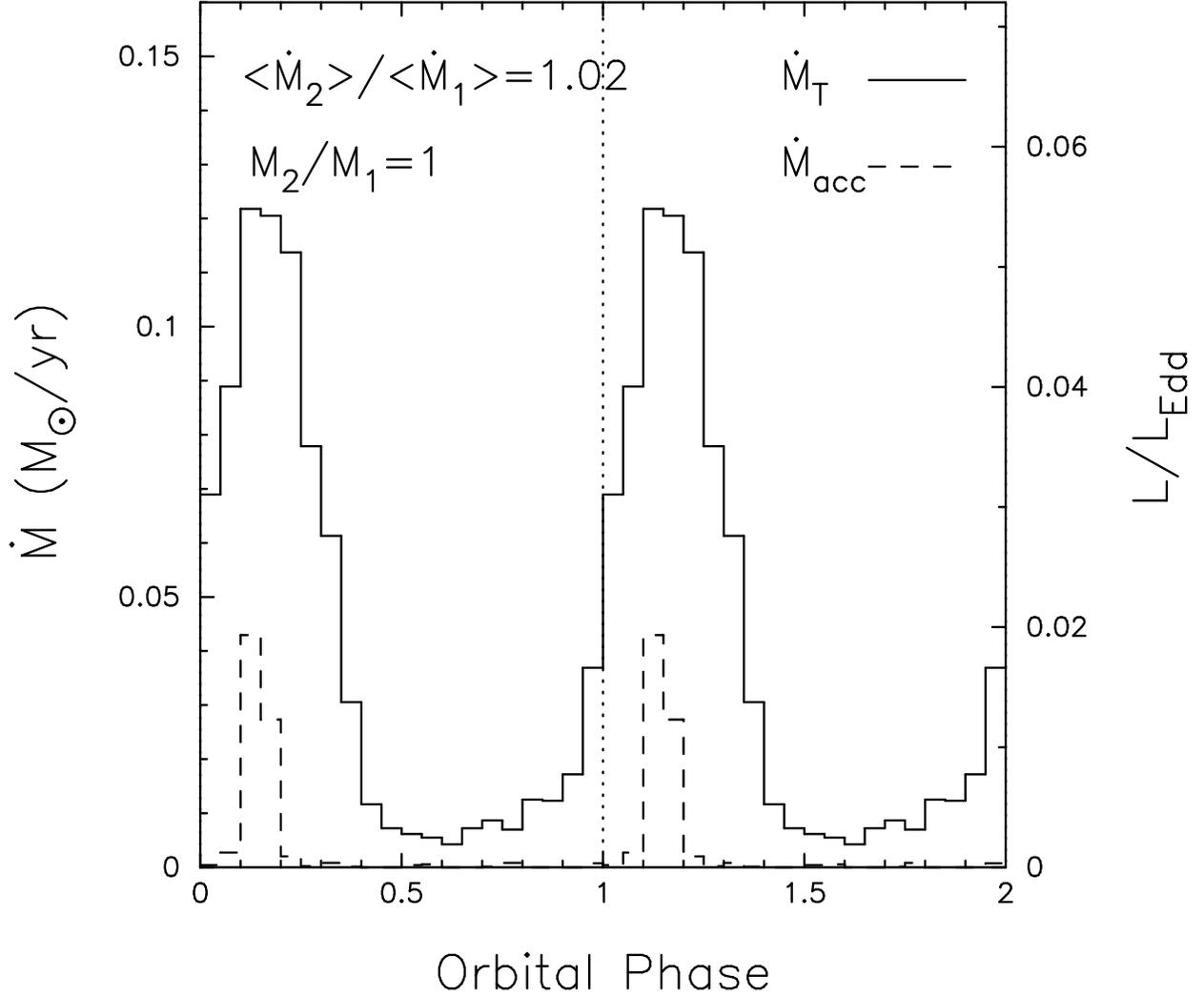}
\caption{
Orbital phase dependence of the mass transfer rate $\dot{M}_{\rm{T}}$ from the 
circumbinary disk to the effective common gravitational radius of the black 
holes, and of the mass accretion rate $\dot{M}_{\rm{acc}}$ at the inner 
simulation boundary, $r_{\rm{in}}=5.0\times10^{-3}a$, of the primary black 
hole.  The right axis shows the bolometric luminosity $L_{\rm{bol}}$  
corresponding to the mass transfer and mass accretion rate with an energy 
conversion efficiency $\eta=0.1$, normalized by the Eddington luminosity 
$L_{\rm{Edd}}$ for a total black hole mass 
$M_{\rm{BH}}=1.0\times10^{8}~M_{\odot}$, where $\eta$ is defined by 
$L_{\rm{bol}}=\eta \dot{M}_{\rm{BH}}c^{2}$.
}
\label{fig:mtr}
\end{figure*}

\begin{figure*}
\resizebox{\hsize}{!}{
\includegraphics*[width=86mm]{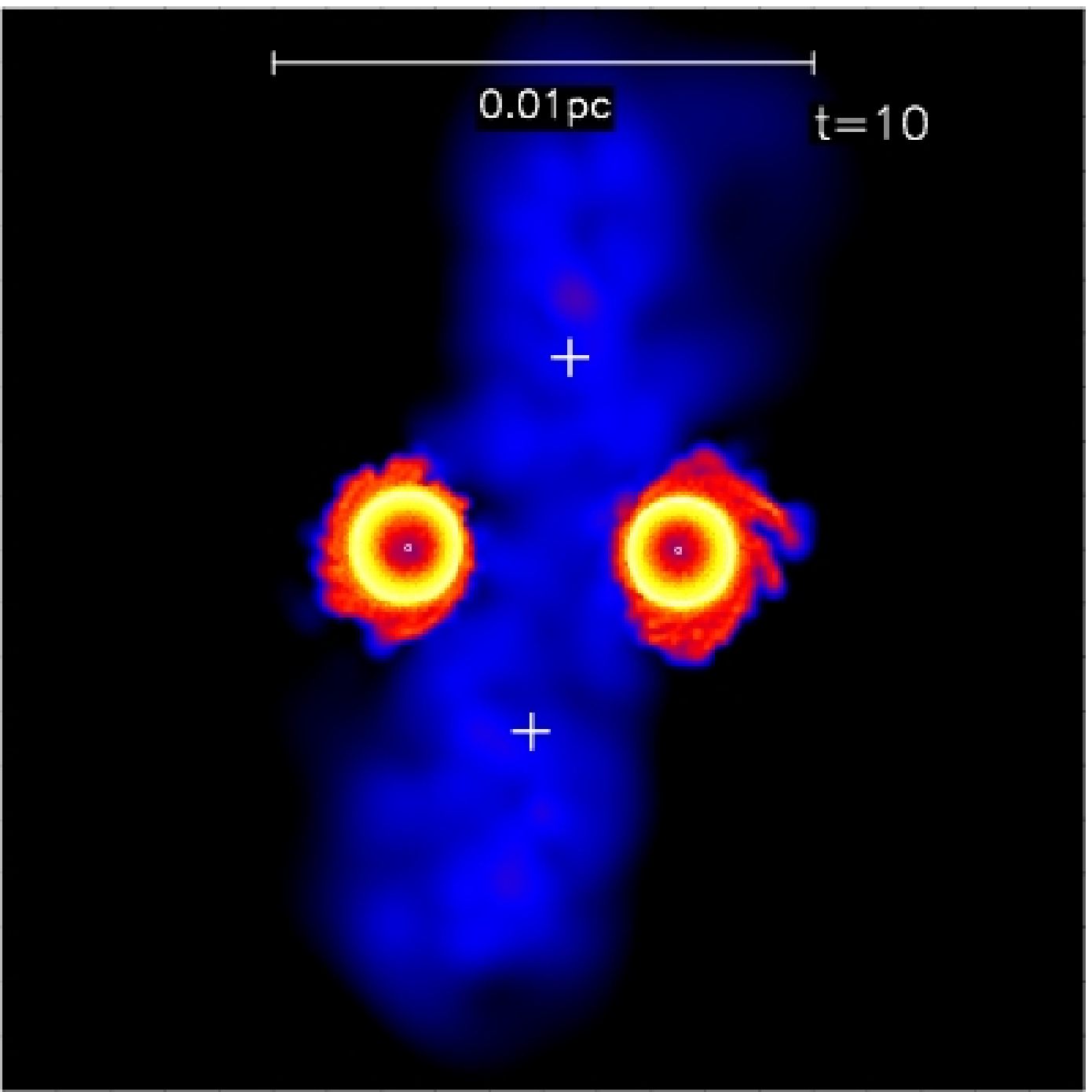}
\includegraphics*[width=86mm]{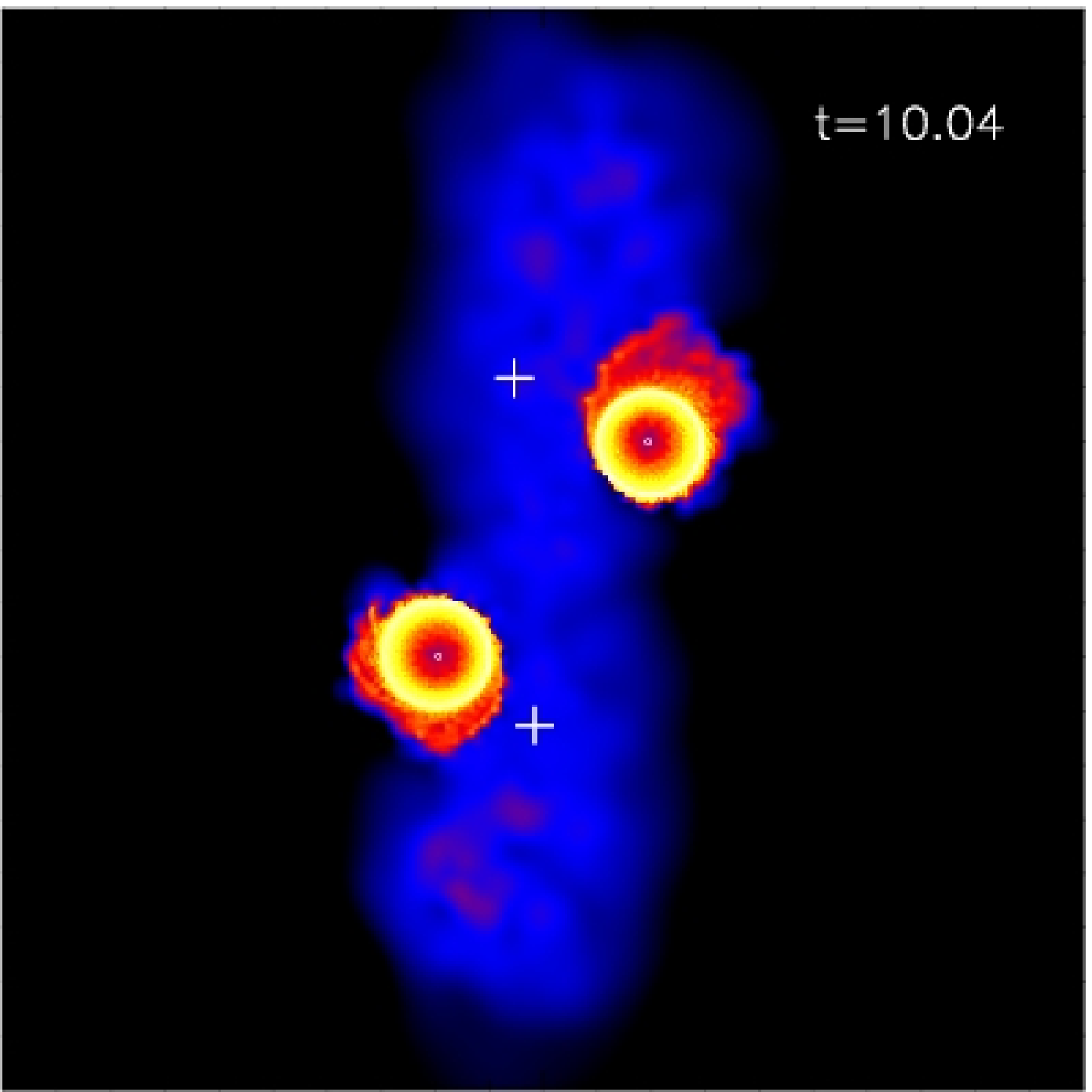}
\includegraphics*[width=86mm]{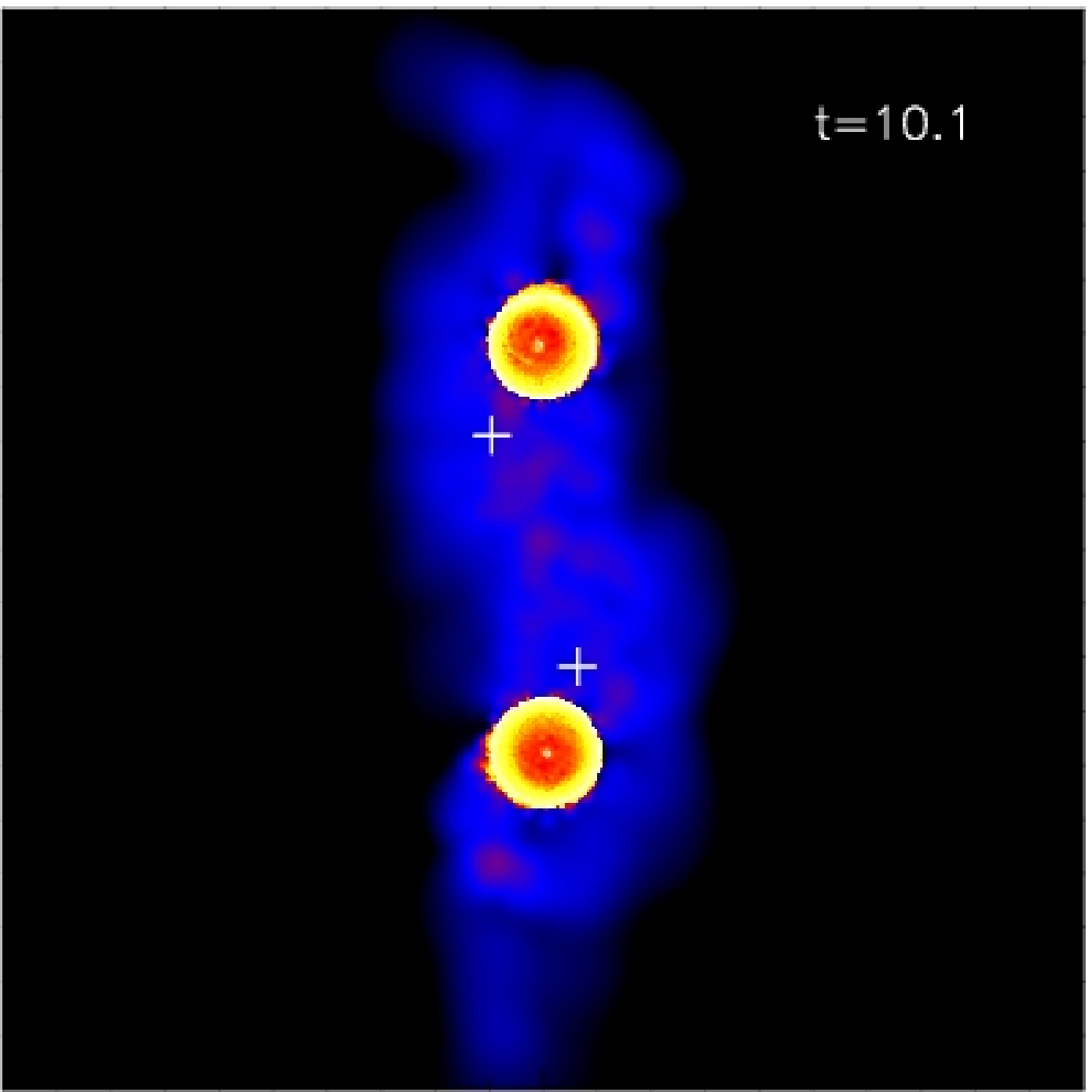}
}
\resizebox{\hsize}{!}{
\includegraphics*[width=86mm]{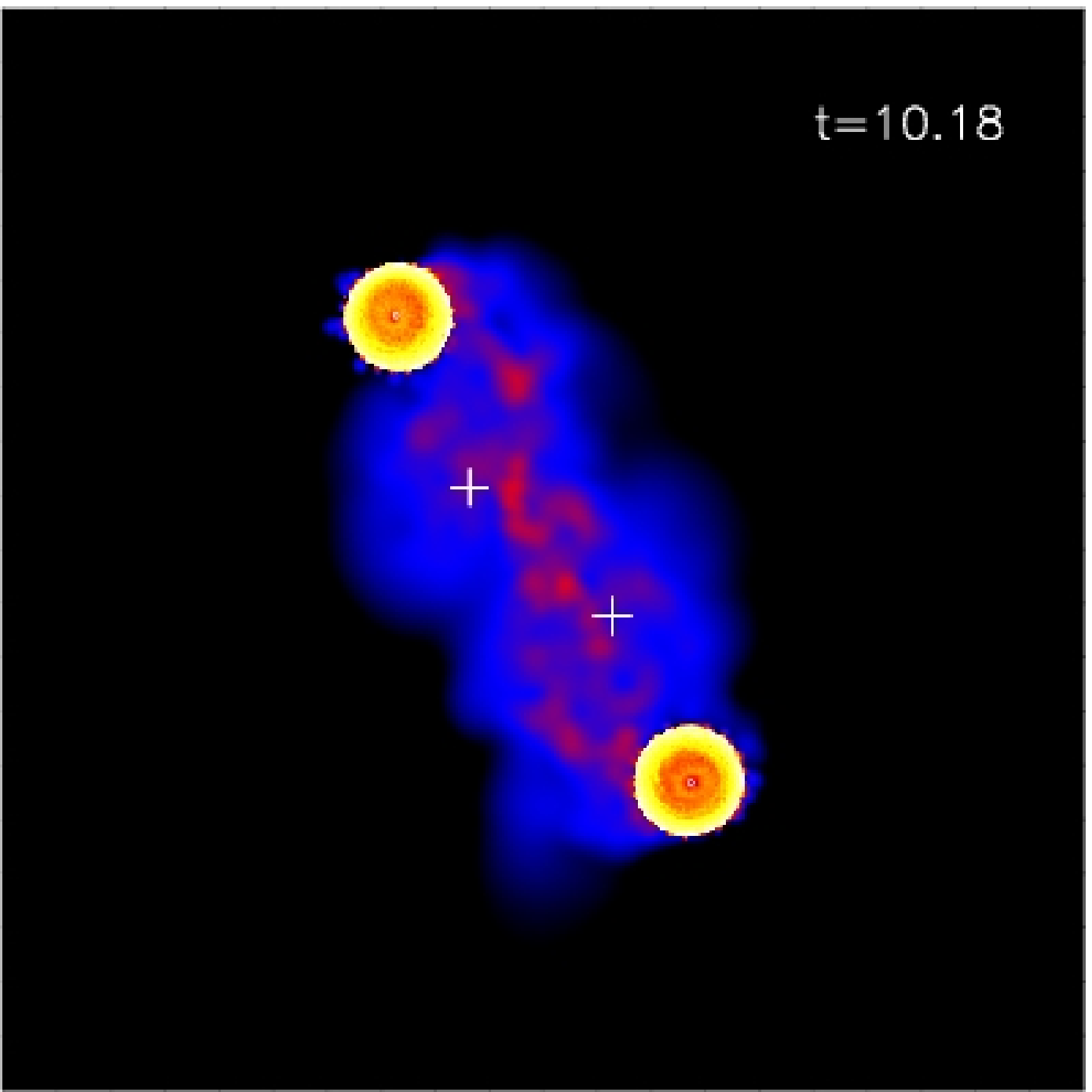}
\includegraphics*[width=86mm]{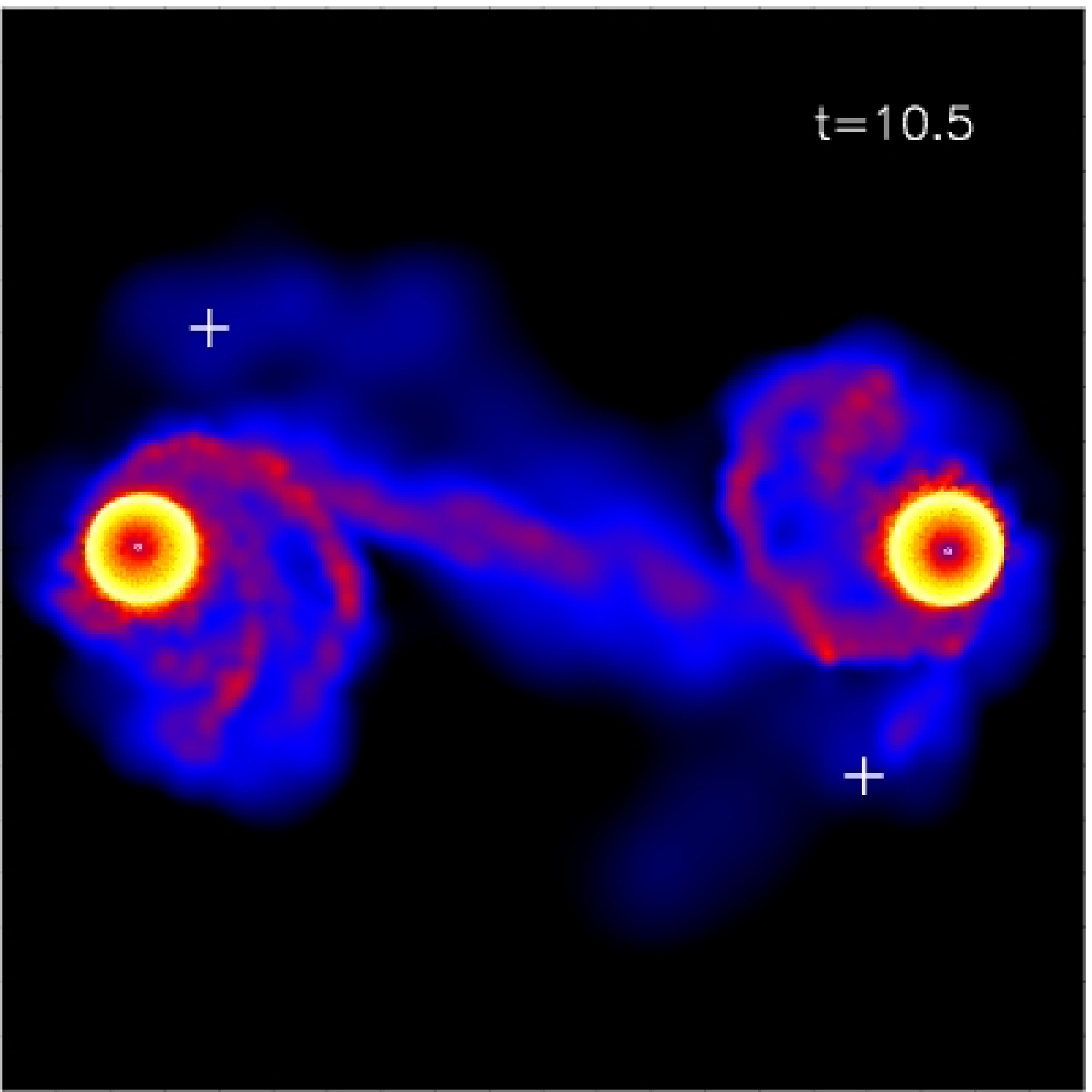}
\includegraphics*[width=86mm]{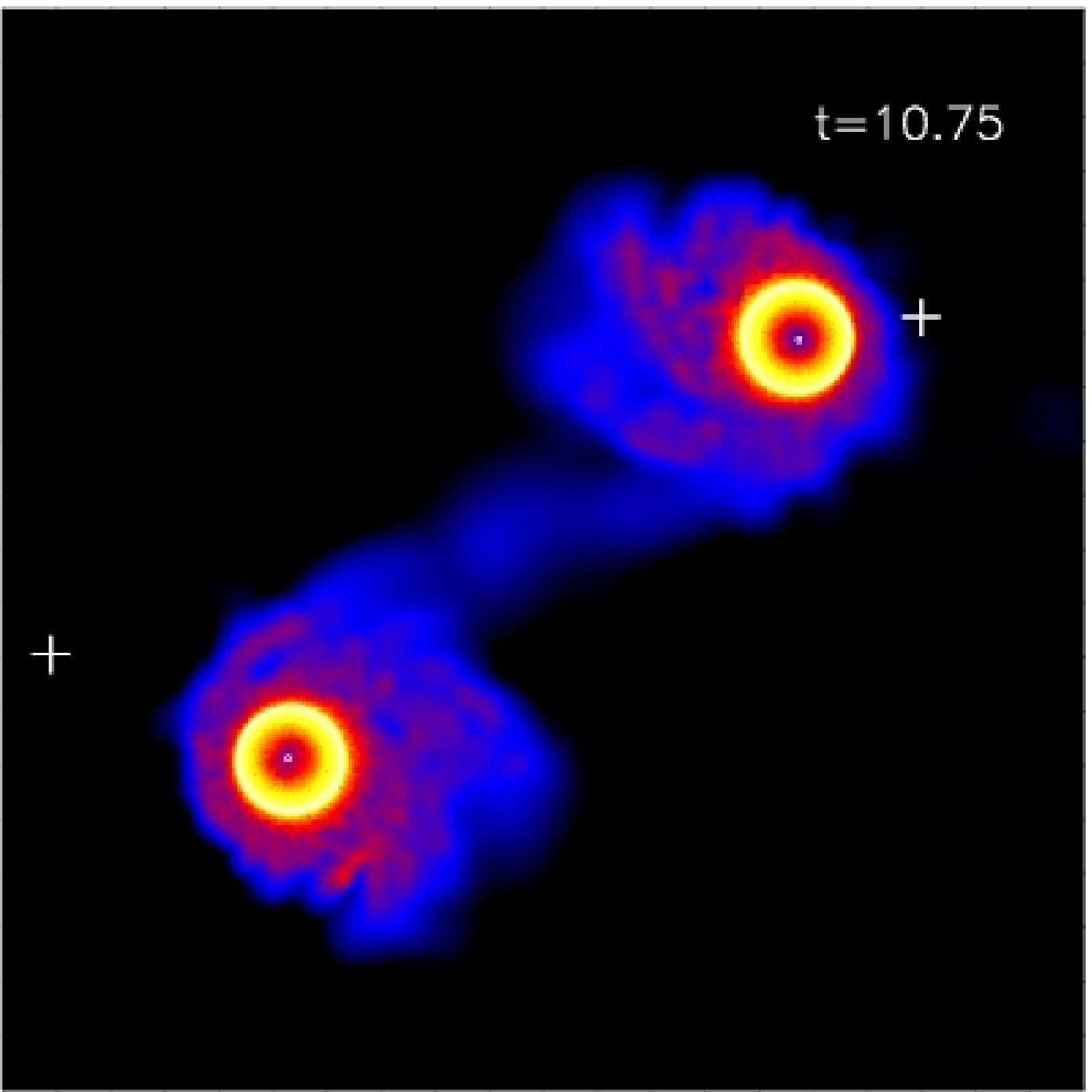}
}
\caption{
Density maps of the two accretion disks around the supermassive BBH rotating 
with $P_{\rm{orb}}\sim9.4\,\rm{yr}$ and $e=0.5$, for the period $10\le{t}\le11$.
Each panel shows on a logarithmic scale the surface density contours over a 
range of 5 orders of magnitude.  The white crosses indicate the positions of 
mass input (see text).  The supermassive {BBH is} rotating in a counterclockwise 
direction.  The time is shown in each panel, and a length scale is shown in 
the top-left panel.  The number of SPH particles during this period ranges 
from $\sim138,000$ to $152,000$.
}
\label{fig:adform}
\end{figure*}
\begin{figure}
\begin{center}
\plottwo{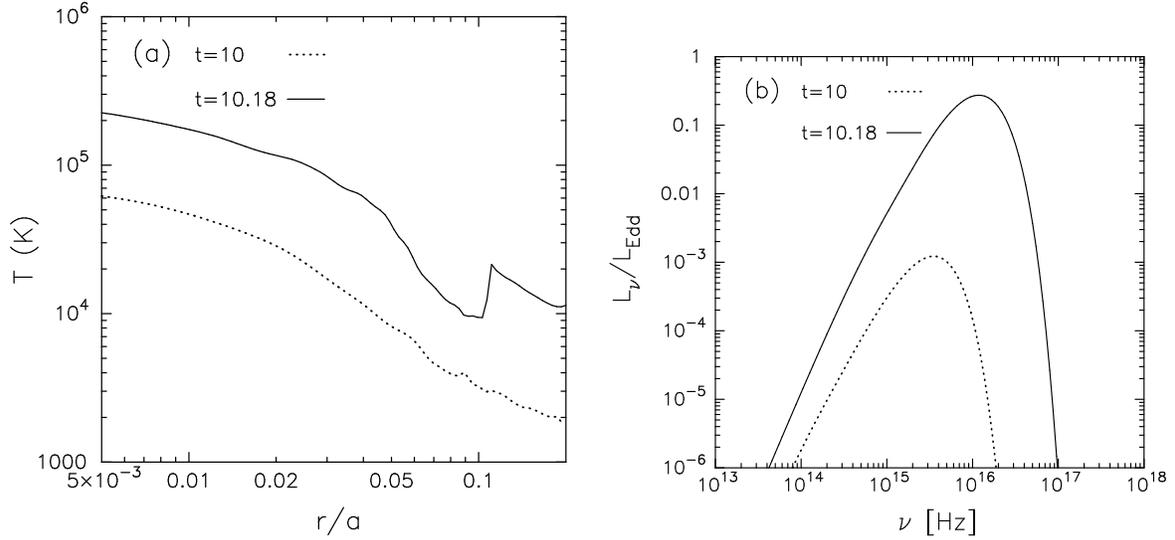}{f3b.eps}
\end{center}
\caption{
({\it a}) Radial distribution of the azimuthally averaged disk temperature
at $t=10.0$ and at $t=10.18$.  ({\it b}) SEDs of the accretion disk around 
the black hole at $t=10.00$ and $t=10.18$.  The integrated luminosity, 
$L_{\nu}={d^{2}}\nu{F_{\nu}}$, is normalized by the Eddington luminosity 
$L_{\rm{Edd}}$ for a total black hole mass 
$M_{\rm{BH}}=1.0\times10^{8}~M_{\odot}$; $F_{\nu}$ is the flux at frequency 
$\nu$ and $d$ is an arbitrary distance from the observer.
}
\label{fig:tksed}
\end{figure}

\begin{figure*}
\plottwo{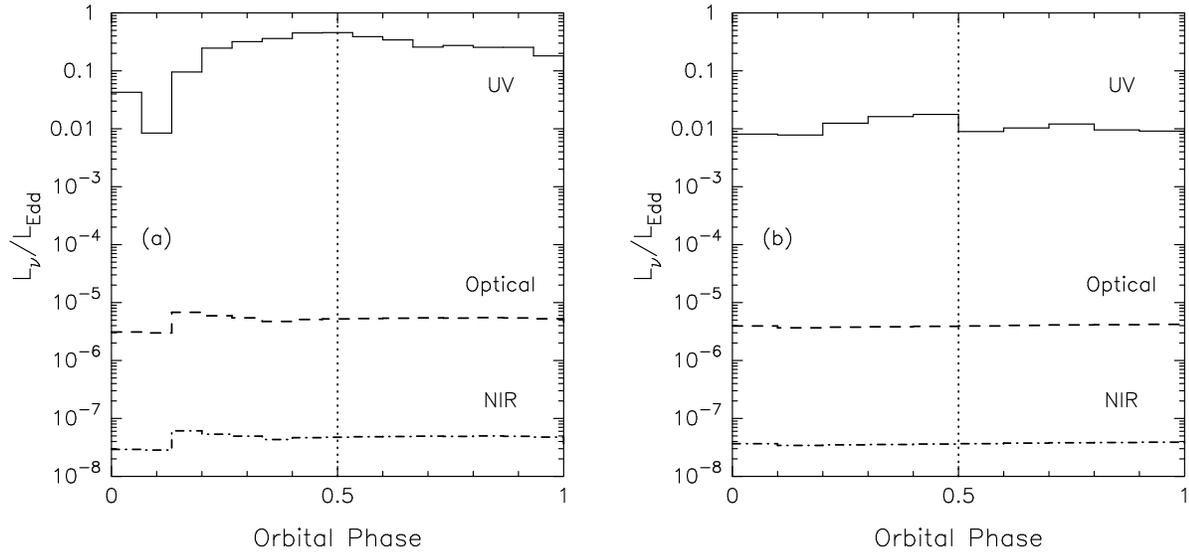}{f4b.eps}
\caption{
Orbital phase dependence of the UV ($\nu\simeq1.0\times10^{16}\,  \rm{Hz}$), 
optical ($B$ band, $\nu \simeq 6.8\times10^{14} \, \rm{Hz}$), and NIR 
($K$ band, $\nu \simeq 1.36\times10^{13} \, \rm{Hz}$) light curves emitted 
from the disk.  Panel ({\it a}) shows the light curves from the simulation 
with mass transfer, and panel ({\it b}) shows the light curves from the 
simulation with no mass transfer.
}
\label{fig:lv}
\end{figure*}


\begin{thebibliography}{}
\bibitem[Armitage \& Natarajan(2002)]{armi1}
Armitage,~P.J., \& Natarajan,~P. 2002, \apj, 567, L9
\bibitem[Armitage \& Natarajan(2005)]{armi2}
-----. 2005, \apj, 634, 921
\bibitem[Artymowicz \& Lubow(1994)]{al94}
Artymowicz, P., \& Lubow, S. H. 1994, \apj, 421, 651
\bibitem[Bate(1995)]{bate295}
Bate, M. R. 1995, PhD thesis, Univ. Cambridge
\bibitem[Bate et al.(1995)]{bate95}
Bate, M. R., Bonnel, I. A., \& Price, N. M. 1995, \mnras, 277, 362
\bibitem[Begelman et al.(1980)]{bege80}
Begelman, M. C., Blandford, R. D., \& Rees, M. J. 1980, Nature, 287, 307
\bibitem[Benz(1990)]{benz90b}
Benz, W. 1990, in Numerical Modeling of Nonlinear Stellar Pulsations: Problems 
and Prospects, ed. R. J. Buchler (Dordrecht: Kluwer), 269 
\bibitem[Benz et al.(1990)]{benz90a}
Benz, W., Bowers, R. L., Cameron, A. G. W., \& Press, W. H. 1990, \apj, 348, 647
\bibitem[Bogdanovi\'c et al.(2008)]{bog08}
Bogdanovi\'c, T., Smith, B. D., Sigurdsson, S., \& Eracleous, M. 2008, \apj, 
submitted (arXiv:0708.0414)
\bibitem[Di Matteo et al.(2005)]{dim05}
Di Matteo, T., Springel, V., \& Hernquist, L. 2005, Nature, 433, 604
\bibitem[Ebisuzaki et al.(1991)]{ebisu91}
Ebisuzaki, T., Makino, J., \& Okumura, S.K. 1991, Nature, 354, 6350
\bibitem[Ferrarese \& Merritt(2000)]{fm00}
Ferrarese, L., \& Merritt, D. 2000, \apj,  539, L9
\bibitem[Gebhardt et al.(2000)]{geb00}
Gebhardt, K., et al. 2000, \apj, 539, L13
\bibitem[Gould \& Rix(2000)]{go00}
Gould, A., \& Rix, H.-H. 2000, \apj, 532, L29
\bibitem[Hayasaki et al.(2007)]{haya07}
Hayasaki, K., Mineshige, S., \& Sudou, H. 2007, \pasj, 59, 427
\bibitem[Hayasaki \& Okazaki(2004)]{haya04}
Hayasaki, K., \& Okazaki, A. T. 2004,  \mnras, 350, 971
\bibitem[Hayasaki \& Okazaki(2005)]{haya05}
------. 2005,  \mnras, 360, L15
\bibitem[Ivanov et al.(1999)]{iv99}
Ivanov, P. B., Papaloizou, J. C. B., \& Polnarev, A. G. 1999, \mnras, 307, 79
\bibitem[Kato et al.(1998)]{kato9807}
Kato,~S., Fukue,~J., \& Mineshige,~S. 1998, Black-Hole Accretion Disks (Kyoto: 
Kyoto Univ. Press)
\bibitem[Komossa et al.(2003)]{komo}
Komossa, S., Burwitz, V., Hasinger, G., Predehl, P., Kaastra, J. S., \& Ikebe, 
Y. 2003, \apj, 582, L15
\bibitem[Lobanov \& Roland(2005)]{lr05}
Lobanov,~A. P., \& Roland,~J. 2005, \aap, 431, 831
\bibitem[Magorrian et al.(1998)]{mag98}
Magorrian, J., et al. 1998, \aj, 115, 2285
\bibitem[Makino(1997)]{makino97}
Makino, J. 1997, \apj, 478, 58
\bibitem[Mayer et al.(2007)]{may07}
Mayer, L., Kazantzidis, S., Madau, P., Colpi, M., Quinn, T., \& Wadsley, J. 
2007, Science, 316, 1874
\bibitem[Merritt \& Ekers(2002)]{me02}
Merritt,~D., \& Ekers,~R. D. 2002, Science, 297, 1310
\bibitem[Milosavljevi\'c \& Merritt(2001)]{milo01}
Milosavljevi\'c, M., \& Merritt, D. 2001, \apj, 563, 34
\bibitem[Monaghan \& Gingold(1983)]{mona83}
Monaghan J. J., \& Gingold, R. A. 1983, J. Comput. Phys., 52, 374
\bibitem[Peters(1964)]{peters64}
Peters, P. C. 1964, Physical Review, 136, 1224
\bibitem[Rodriguez et al.(2006)]{rod06}
Rodriguez, C., Taylor, G. B., Zavala, R. T., Peck, A. B., Pollack, L. K., \& 
Romani, R. W. 2006, \apj, 646, 49
\bibitem[Roos(1981)]{roos81}
Roos, N. 1981, \aap, 104, 218
\bibitem[Shakura \& Sunyaev(1973)]{ss73}
Shakura, N. I., \& Sunyaev, R. A. 1973, \aap, 24, 337
\bibitem[Sillanp\"a\"a et al.(1988)]{sill88}
Sillanp\"a\"a, A., Haarala, S., Valtonen, M., Sundelius, B., \& Byrd, G. G. 
1988, \apj, 325, 628
\bibitem[Sudou et al.(2003)]{sudou}
Sudou, H., Iguchi, S., Muratai, Y., \& Taniguchi, T. 2003, Science, 300, 1263
\bibitem[Sundelius et al.(1997)]{sun97}
Sundelius, B., Wahde, M., Lehto, H. J., \& Valtonen, M. J. 1997, \apj, 484, 180
\bibitem[Yu(2001)]{yu01}
Yu, Q. 2001, \aap, 377, 17 
\bibitem[Yu(2002)]{yu02}
------. 2002, \mnras, 331, 935
\bibitem[Yu \& Tremaine(2002)]{yutre02}
Yu, Q., \& Tremaine, S. 2002, \mnras, 335, 965

\end{thebibliography}
\end{document}